\documentclass[aps,prd,twocolumn,showpacs,natbib]{revtex4-1}

\usepackage{dcolumn}
\usepackage{bm}

\usepackage{times} 
\usepackage{graphicx,color}
\usepackage{amssymb,amsmath}

\mathchardef\bigtilde="0365

\def\la{\langle}
\def\ra{\rangle}
\def\s{\sigma}

\begin{document}

\title{ Inverse Laplace transform on the lattice spacing}
\author{Hirofumi Yamada}
\email{yamada.hirofumi@it-chiba.ac.jp}
\affiliation{
Division of Mathematics and Science, Chiba Institute of Technology, 
\\Shibazono 2-1-1, Narashino, Chiba 275-0023, Japan}

\date{\today}

\begin{abstract}
{Inverse Laplace transform on the lattice spacing is introduced as a computational framework of the extrapolation of the strong coupling expansion to the scaling region.   We apply the transform to the two-dimensional nonlinear $O(N)$ model at $N\ge 3$ and show that the approximation of the continuum limit of the susceptibility agrees with the existing theoretical and Monte Carlo data.}
\end{abstract}


\maketitle

The lattice spacing $a$ serves as the fundamental parameter in lattice formulation of physical models.  In addition to the role of ultraviolet regulator, it controls almost all physical quantities, since the bare coupling $g$ depends on it due to the dimensional transmutation.   Further it is a direct measure of how the model being far from or close to the continuum limit.   
As an attempt to explore the theoretical framework of approaching to the continuum limit from the view point of the expansion around $a=\infty$, we have studied the delta expansion and Pad\'e-Borel technique in recent works \cite{yam,yam1}.    

We are now aware that the techniques used in \cite{yam,yam1} can be focused to the inverse Laplace transform on the lattice spacing $a$, which may be extended to complex values for the sake of the transformation.  In this Letter, we give the basis of the method and, as an illustration, revisit the two dimensional (2D) $O(N)$ sigma model at $N\ge 3$ to demonstrate the successful approximation of the susceptibility in the continuum limit. 

As a conventional manner, it is convenient and natural to use $a$ in dimensionless combination with other dimensional parameter.  The appropriate one would be the scale $\Lambda_{L}$ but it is hidden with $a$ inside the bare coupling.  In a sense, the mass gap or the correlation length is a concrete realization of the scale $\Lambda_{L}$.   They can be computed in the strong coupling expansion and the result fixes the functional relation between them and $g$ at large $a$.   Thus, for example, the mass gap (or square in some models) combined with $a$ to lose mass dimension is an appropriate parameter replacing $\Lambda_{L}a$.  Let us denote such a variable as $M$.   

Now we introduce the inverse Laplace transform as follows:  For a quantity $Q(M)$ given as a function of $M$, we make complex extension of $M$ to $z\in {\mathbb C}$ by adding pure imaginary part to $M$.  Then, consider the superposition over the imaginary part and define inverse Laplace transform for $t\in {\mathbb R}$ by
\begin{equation}
\bar Q(t)=\int_{M-i\infty}^{M+i\infty}\frac{dz}{2\pi i}\frac{\exp(tz)}{z}Q(z):=L^{-1}[Q].
\label{borel}
\end{equation}
The contour of integration is parallel to the imaginary axis.  
 Laplace transform recovers the original function by
\begin{equation}
Q(z)=z\int_{0}^{\infty}dt \exp(-tz)\bar Q(t):=L[\bar Q].
\label{lap}
\end{equation}
As long as the lattice models under consideration are subject merely to second or higher order transitions, there is no singularity at positive real part of $z$ axis.   Here, we further assume that $Q(z)$ has no singularity in the right-half plane of $z$.  This ensures that the contour position specified by $M$ is arbitrary as long as $M>0$ and $M$ disappears in the result.   The first singularity to be met with would be, when we move the contour to the left, found  at $z=0$ if the singularity coming from $z^{-1}$ in the measure is not removed.   

As in the Fourier transform, the scaling behavior of $Q(M)$ is connected to the large $t$ behavior of $\bar Q(t)$ (see (\ref{borel}) or (\ref{lap})).   
In turn, small $t$ behavior is related to the large $a$ behavior of $Q$.  Then,  it is apparent from the following basic result,
\begin{equation}
L^{-1}[z^{-n}]=\frac{1}{n!}t^n,
\label{fact}
\end{equation}
that the series of $\bar Q$ in $t$ is usually an entire one and quite adequate to approximate the scaling behavior of $Q(M)$.  The physical interpretation of $t$ is still unclear to us.  However, the study of $\bar Q(t)$ in the approximate computation of the continuum limit of $Q(M)$ is much more tractable than $Q$ itself.

Now, we apply our scheme to 2D nonlinear $O(N)$ sigma model at $N\ge 3$ and illustrate how working in the $t$-space endowed with Pad\'e approximation is effective in the study of continuum limit via strong coupling expansion.  
The standard action of the nonlinear sigma model on square lattice reads
\begin{equation}
S=-\beta\sum_{\bf n}\sum_{\mu=1,2}\vec{\s}_{\bf n}\cdot\vec{\s}_{\bf n+\bf e_{\mu}},
\label{sigmaaction}
\end{equation}
where ${\bf e}_{1}=(1,0),\,{\bf e}_{2}=(0,1)$ and $\beta$ is the reciprocal of the naive bare coupling $g$,
\begin{equation}
\beta=1/g.
\end{equation}
The vector $\vec{\sigma}=(\sigma_{1},\sigma_{2},\cdots, \sigma_{N})$ is constrained to satisfy $\vec{\s}^2=N$ at all sites.  We define the basic parameter $M$ by the zero momentum limit of two-point function of $\sigma$ (see \cite{yam}).  It was obtainable from the susceptibility and second moment both of which were computed in $\beta$ up to $\beta^{21}$ \cite{butera}.  For our approach, $\beta$ must be rewritten in $M$.  It is simply given by the inversion giving \cite{yam}
\begin{equation}
\beta=\frac{1}{M}- \frac{4}{M^2}+\frac{2 (10 N+19)}{(N+2)M^3}-\frac{8 (14N+25)}{(N+2)M^4}+\cdots.
\label{strong1}
\end{equation}

At small $M$ near the continuum limit, we notice the perturbative result of correlation length $\xi$ \cite{fal,col,col2,shin},
\begin{eqnarray}
\xi&=& C_{\xi} \exp\Big[\frac{\beta}{-b_{0}}\Big]\Big(\frac{\beta}{-b_{0}}\Big)^{\frac{-1}{N-2}}\Big(1+\sum_{k=1}^{\infty}\frac{b_{k+1}}{\beta^k}\Big),\label{xirg}\\
b_{0}&=&-\frac{N-2}{2\pi N},\quad b_{1}=-\frac{N-2}{(2\pi N)^2},\nonumber\\
b_{2}&=&\frac{1}{N(N-2)}(-0.0490-0.0141N),\nonumber\\
b_{3}&=&\frac{1}{N^2(N-2)^2}\nonumber\\
& &\times(0.0444+0.0216N+0.0045N^2-0.0129N^3).\nonumber
\end{eqnarray}
The mass square $M$ is the momentum counter part of $\xi^2$ and its dependence on $\beta$ should be identical with (\ref{xirg}) but possibly accompanied by another multiplicative constant, say $C_{M}$.   However, Monte Carlo data \cite{ed} showed that the difference is less than a percent at $N=4$.   Since the two constants agree with each other in the large $N$ limit due to the Gaussian nature of two point function, the difference between $C_{\xi}$ and $C_{M}$ may actually be negligible for all $N\ge 3$.   Thus, we neglect the slight difference of  $C_{M}$ and $C_{\xi}$ and simply substitute $M^{-1/2}$ into $\xi$ in (\ref{xirg}), which is valid as long as $\beta$ is large enough.  
The constant $C_{\xi}$ is specified only non-perturbatively and was obtained by Hasenfratz, Maggiore and Niedermayer \cite{hasen} as
\begin{equation}
C_{\xi}=32^{-1/2}\Big(\frac{e^{1-\pi/2}}{8}\Big)^{\frac{1}{N-2}}\Gamma\Big(1+\frac{1}{N-2}\Big).
\label{cxi}
\end{equation}
By solving (\ref{xirg}) for $\beta$, we have
\begin{equation}
\beta\sim \frac{N-2}{4\pi N}\log x+\frac{1}{2\pi N}\log\big(\frac{1}{2}\log x\big)+O(\frac{\log\log x}{\log x}),
\label{weakbeta}
\end{equation}
where
\begin{equation}
x=(M C_{\xi}^2)^{-1}
\end{equation}
Since the complete expression up to the four loop level and its transform are lengthy, we refer them in \cite{yam}.   The expansion (\ref{weakbeta}) will be used when physical quantities are expressed in $M$ in the weak coupling region.
 
Representation of bare coupling in $t$-space is adequate for the extrapolation of the strong coupling series (\ref{strong1}) to the scaling region.  The inverse Laplace transform is obtained by changing $M^{-n}$ to $t^n/n!$, giving
\begin{equation}
\bar\beta=t- \frac{4}{2!}t^2+\frac{2 (10 N+19)}{(N+2)3!}t^3-\frac{8 (14N+25)}{(N+2)4!}t^4+\cdots.
\label{strong1borel}
\end{equation}     
Using the transformed bare coupling (\ref{strong1borel}), one can confirm at least around or above $N=6$, the continuum scaling of $\bar\beta$, the inverse Laplace transform of (\ref{weakbeta}).  Then, the mass gap in the continuum limit can be estimated by fitting the four loop $\bar\beta_{pert}$ to (\ref{strong1borel}).  The process and the result is exactly the same as that presented in \cite{yam}.  Hence we suffice ourselves to say the conclusion that the scaling behavior was observed in $\bar\beta(t)$ in series of $t$ and the estimation of $C_{\xi}$ agreed with the result (\ref{cxi}).

Now, we present a new attempt to compute the susceptibility $\chi$ near the continuum limit via its strong coupling expansion.   Though the problem was discussed in \cite{yam2} according to the technique of Pad\'e-Borel approximants, $\beta$ was taken as the function of $\chi$.  The proposal in this brief report is to take a unified approach in the study of the continuum limit by choosing $\Lambda_{L}a$ or $M$ as the fundamental parameter.   Here, we regard $\chi$ as the function of $M$ and consider the inverse Laplace transform of $\chi$ on complex $M$ plane.

The susceptibility $\chi$ is defined by
\begin{equation}
\chi=\frac{\beta}{N}\sum_{\bf x}\la\vec{ \sigma}_{\bf 0}\cdot \vec{ \sigma}_{\bf x}\ra.
\end{equation}
From perturbative renormalization group at $N\ge 3$, $\chi$ near the continuum limit behaves as
\begin{equation}
\chi\sim \beta C_{\chi}\exp\Big[\frac{2\beta}{-b_{0}}\Big]\Big(\frac{\beta}{-b_{0}}\Big)^{\eta} \Big(1+\frac{c_{2}}{\beta}+\frac{c_{3}}{\beta^2}+\cdots\Big),
\label{chi}
\end{equation}
where $\eta=2b_{1}/b_{0}^2+g_{0}/b_{0}$ and  \cite{fal,col,col2,shin}
\begin{eqnarray}
g_{0}&=&\frac{N-1}{2\pi N},\nonumber\\
c_{2}&=&\frac{1}{N(N-2)}(-0.1888+0.0626N),\nonumber\\
c_{3}&=&\frac{1}{N^2(N-2)^2}\nonumber\\
& &\times(0.1316+0.0187N-0.0202N^2-0.0108N^3).\nonumber
\end{eqnarray}
The constant $C_{\chi}$ in (\ref{chi}) is of non-perturbative and not analytically known.   We estimate  $C_{\chi}$ in later.

First of all, we express $\chi$ in terms of the mass square $M$ in both regions at $\beta\gg 1$ and $\beta\ll 1$.  At weak coupling, using the perturbative result (\ref{weakbeta}) and (\ref{chi}), we find
\begin{widetext}
\begin{eqnarray}
\chi&=&\frac{C_{\chi}(N-2)}{2\pi N}x\Big(\frac{\log x}{2}\Big)^{-1/(N-2)}\Big[1+\frac{2(a_{1}-\log\log x)}{(N-2)^2\log x}+\frac{a_{2}+a_{3}\log\log x+a_{4}(\log\log x)^2}{(N-2)^4(\log x)^2}\Big],\\
a_{1}&=&-2(N-2)\pi(2N b_{2}-c_{2})+\log 2,\nonumber\\
a_{2}&=&8(N-2)^2\pi[N(b_{2}+6N\pi b_{2}^2-4N\pi b_{3}-4\pi b_{2}c_{2})+2\pi c_{3}]-4(N-2)(-1+2(N-1)\pi(2N b_{2}-c_{2}))\log 2\nonumber\\\
& &+2(N-1)(\log 2)^2,\nonumber\\
a_{3}&=&4(N-2)(-1+2(N-1)\pi(2N b_{2}-c_{2}))-4(N-1)\log 2,\nonumber\\\
a_{4}&=&2(N-1).\nonumber
\label{chi1}
\end{eqnarray}
\end{widetext}
We need to compute inverse Laplace transform of $z^{-1}(\log z^{-1})^{-\alpha}$ $(\alpha>0$ and $z$ stands the complex extension of $M$).  We note that there is a singularity at $z=1$ coming from $(\log z)^{-\alpha}$.    This is an artifact of perturbation but as a consequence we need another cut in the $z$ plane at $[1, \infty)$.   Thus we restrict the location of the contour in the contour integral
\begin{equation}
L^{-1}[z^{-1}(\log z^{-1})^{-\alpha}]=\int_{M-i\infty}^{M+i\infty}\frac{dz}{2\pi i}\frac{e^{tz}}{z}z^{-1}(\log z^{-1})^{-\alpha},\nonumber
\end{equation}
to be included within the strip $0<\Re[z]<1$.  
The perturbative region corresponds to the region $t\gg 1$ and the asymptotic expansion of the integral reads
\begin{widetext}
\begin{equation}
L^{-1}[z^{-1}(\log z^{-1})^{-\alpha}]
=t(\log t+\gamma_{E})^{-\alpha} 
\Big(1+\frac{\alpha}{\log t+\gamma_{E}}+\frac{\alpha(\alpha+1)}{(\log t+\gamma_{E})^2}(1-\zeta(2)/2)+\cdots\Big).
\label{result}
\end{equation}
The transform of terms of the type $z(\log z)^{-\alpha}(\log\log z)^{k}$ $(k=1,2)$ can be readily obtained from (\ref{result}) by shifting $\alpha\to \alpha+\epsilon$ and then expanding the result in $\epsilon$.  In this manner we arrive at
\begin{eqnarray}
\bar\chi&=&\frac{C_{\chi}(N-2)}{2\pi N}t\Big(\frac{\log T}{2}\Big)^{-1/(N-2)}\bigg[1+\frac{1}{(N-2)\log T}+\frac{(N-1)(1-\zeta(2)/2)}{(N-2)^2(\log T)^2}\nonumber\\
& &+\frac{2 a_{1}(1+\frac{N-1}{(N-2)\log T})-2(-\frac{1}{\log T}+\log\log T(1+\frac{N-1}{(N-2)\log T}))}{(N-2)^2\log T}+\frac{a_{2}+a_{3}\log\log T+a_{4}(\log\log T)^2}{(N-2)^4(\log T)^2}+\cdots\bigg]\nonumber\\
&:=&\bar \chi_{pert},
\label{bchipert}
\end{eqnarray}
\end{widetext}
where
\begin{equation}
T=t/ C_{\xi}^2 e^{\gamma_{E}}.
\end{equation}

At strong coupling, the susceptibility is obtainable in powers of $\beta$.   The result of Butera and Comi \cite{butera} gives us
$
\chi=\beta (1+4\beta+12\beta^2+\frac{72+32N}{N+2}\beta^3+\cdots).
$
Then, substituting (\ref{strong1}) into the series, we obtain the series of $\chi$ expanded in $1/M$,
\begin{equation}
\chi=\frac{1}{M}-\frac{2}{(N+2)M^2}+\frac{2(59N^2+364N+480)}{(N+2)^2(N+4)M^3}+\cdots.
\end{equation}  
The inverse Laplace transform is readily obtained by changing $M^{-n}\to t^{n}/n!$.  The result reads
\begin{eqnarray}
\bar \chi&=&t-\frac{2t^2}{2!(N+2)}+\frac{2(59N^2+364N+480)t^3}{3!(N+2)^2(N+4)}+\cdots\nonumber\\
&:=&\bar\chi_{str}.
\label{chiborel}
\end{eqnarray}

\begin{figure}[h]
\centering
\includegraphics[scale=0.6]{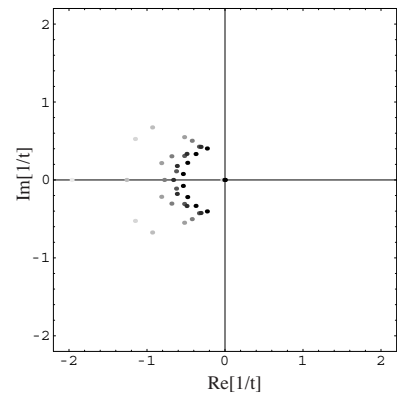}
\label{poles}
\caption{Pole distribution of diagonal $\chi_{str}$ for  $N=3$ at orders $[3/3]$, $[4/4]$, $\cdots$ and $[9/9]$.  The gray level of dots represents the order.  Lightest ones correspond to poles at $6$th and black ones at $18$th.}
\end{figure}

To study the scaling behavior of $\bar \chi$ from the series (\ref{chiborel}), the series expansion of $\bar \chi$ is insufficient for accurate results, though inverse Laplace transform substantially enlarged the effective region of the truncated series.  It requires expansion up to quite large orders which no one performed yet.  Fortunately the status of approximation task is improved by the use of Pad\'e approximants.  The Pad\'e approximant method replaces the power series by the ratio of polynomials in $t$.  The degrees $m$ and $n$ of the numerator and the denominator must have sum which agrees with the truncation order of $\bar\chi_{str}$.  Since, at large $t$, it is perturbatively known that $\bar\chi$ behaves linearly in $t$ with the logarithmic correction, we confine ourselves with the rational functions at diagonal and near-diagonal entries in the Pad\'e table.

Denoting the rational approximant as $[m/n]$, we have examined the pole distributions of $[n/n]$, $[n+1/n-1]$, $[n-1/n+1]$ and $[n+1/n]$ cases.   We found that $[n/n]$ and $[n+1/n-1]$ approximants have no poles in the right-half plane of $t^{-1}$ for all $N\ge 3$.  Further, the pole distribution from $4$th to $18$th orders at $N=3$ shows that the poles in the left-half plane move toward the origin as the order increases (see Fig.1).  It is highly conceivable that at the infinite order, the set of infinite poles collapse to the origin from the left.   This is a reflection that the transformed series in $t$ is an entire function 
and the set of degenerate poles represents the perturbative singularity implied by logarithms in (\ref{bchipert}).   On the other hand, $[n-1/n+1]$ and $[n+1/n]$ approximants are plagued with poles in the right-half plane.  Even those poles are merely artifacts of Pad\'e approximant, they badly influence the behavior of $\bar\chi$ on the real $t$-axis.  Though $[n+1/n]$ approximants are the most important for the perturbative behavior of  $\chi$ being linear in $t$,  we therefore discard the result from $[n+1/n]$ ($[n-1/n+1]$ type is also discarded, of course).  
\begin{figure}[h]
\centering
\includegraphics[scale=0.6]{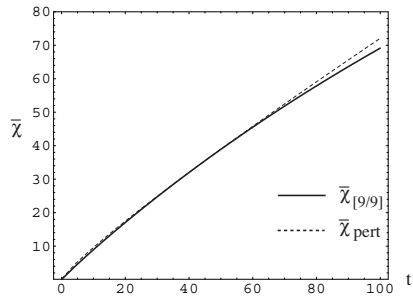}
\label{comparison}
\caption{$\bar\chi_{pert}$ (dotted) and $\bar\chi_{[9/9]}$ for $N=3$.  In plot of $\bar\chi_{pert}$, we used for $C_{\chi}$ the estimated value $C_{\chi}=0.012104$.}
\end{figure}

\begin{figure}[h]
\centering
\includegraphics[scale=0.65]{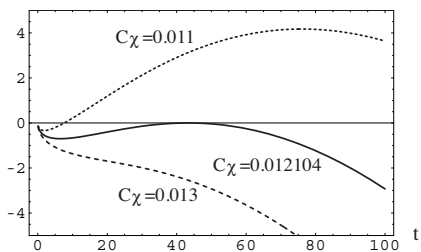}
\label{comparison2}
\caption{Plots of $\bar\chi_{[9/9]}-\bar\chi_{pert}$ at $N=3$ for three values of $C_{\chi}$, $C_{\chi}=0.011$, $0.012104$ and $0.013$.}
\end{figure}

After these observations, we compare Pad\'e approximants at $[n/n]$ and $[n+1/n-1]$ to the $4$-loop scaling behavior $\bar\chi_{pert}$ by using the exact result of $C_{\xi}$ \cite{hasen}.   We have ascertained that Pad\'e approximants approximately recover the scaling behavior for $N\ge 3$.   For example at $N=3$,  from the plot of $\bar\chi_{[9/9]}$ and $\bar\chi_{pert}$ in Fig. 2, we see that the two functions are close with each other for $C_{\chi}=0.012104$ in the interval around $t\sim 40$.   To clarify the effect of $C_{\chi}$ on the behavior, we have plotted in Fig. 3 the difference $\bar\chi_{pert}-\bar\chi_{[9/9]}$ at $C_{\chi}=0.011,\, 0.012104,\, 0.013$.   It is clearly seen that the value $C_{\chi}=0.012104$ achieves the best optimization.  Only at this value, both of $\bar\chi_{pert}$ and $\bar\chi_{[9/9]}$ and their derivatives agree at a point.   In this way, we can estimate unknown non-perturbative constant $C_{\chi}$ for other values of $N$ by using $[m/n]=[9/9]$ and $[10/8]$ forms of approximants.  The result is summarized in Table I.  
\begin{table}[h]
\caption{
Estimated values of the non-perturbative constant $C_{\chi}$.    On estimation we have used Pad\'e-Borel approximants of $\bar\beta$ at order $18$th for the cases $[9/9]$ and $ [10/8]$.     $C_{BT}$ means the estimation due to Butera and Comi \cite{butera}. } 
\label{tab:1}  
\begin{ruledtabular}
\begin{tabular}{cccc}
$ N $ & $ [9/9]$ & $  [10/8]$ &  $  C_{BT}$ \\
\noalign{\smallskip}\hline\noalign{\smallskip}
$ 3$ &   $0.012104$ & $0.012034$ &  $   $  \\ 
$ 4$ & $0.037018$ & $0.036971$   &  $0.034$  \\ 
$5$ & $0.059836 $  & $0.059809$    &   $0.059$ \\ 
$6$ &$0.078058$   & $0.078042$    &   $0.077$ \\ 
$ 7$ & $0.092404 $  & $0.092395$   &  $     $  \\ 
$8$ & $0.103830$ & $0.103824$    & $  0.1035$  \\ 
$ 9$ &$0.113083$  & $0.113079 $    & $       $\\ 
$10$ &$0.120703$ & $0.120700$     &   $0.1212$\\ 
$20$ &$0.157114 $  & $ 0.157113$    &  $       $\\
\end{tabular}
\end{ruledtabular}
\end{table}

For all values of $N$ at which Butera and Comi made theoretical estimation of $C_{\chi}$, our results are in good agreement.   There is also 
Monte Carlo data for $N=3$, $4$ and $8$.  At $N=3$, the two results, $C_{\chi}=0.0146(10)$ \cite{cara} and $C_{\chi}=0.0130(5)$ \cite{all} were reported.  
  At $N=4$, $C_{\chi}=0.0329(16)$ in \cite{ed} and $C_{\chi}=0.0383(10)$ in \cite{cara}.  At $N=8$, $C_{\chi}=0.1037(4)$ in \cite{cara} and $C_{\chi}=0.1028(2)$ in \cite{all}.   Though small discrepancy  remains, our estimates are consistent with these Monte Carlo data.  The values of $[9/9]$ and $ [10/8]$ entries are close with each other for all $N$ and this fact indicates the stability of the approximation.   As a whole, the level of agreement with existing results is improved than the result of \cite{yam2}.
  
As demonstrated by the examination presented so far, inverse Laplace transform on the lattice spacing is found to be effective in the approximation of the continuum limit.  Rather than the real space of $a$, the dual space is convenient to consider scaling properties.  The inverse Laplace transform preserves the essential information at scaling region and creates an entire function in $t$.   It is worth exploring the approach as an alternative route to the continuum limit from the strong coupling expansion.


\begin{thebibliography}{9}

\bibitem{yam} Yamada H  2011 Phys. Rev. D84 105025.
\bibitem{yam1}  Yamada H 2009 J. of Phys. G, 36 025001.
\bibitem{butera} Butera P and Comi M 1996 Phys. Rev. B 54 15828.
\bibitem{fal} Falcioni M and Treves A 1986 Nucl. Phys. B265 671.
\bibitem{col} All\'es B, Caracciolo S, Pelissetto A and Pepe M 1999 Nucl.Phys. B562 581.
\bibitem{col2} Caracciolo S and Pelissetto A 1995 Nucl.Phys. B455 619.
\bibitem{shin} Shin D 1999 Nucl.Phys. B546 669.
\bibitem{ed} Edwards R G, Ferreira E Goodman J and Sokal A D 1992 Nucl. Phys. B380 621.
\bibitem{hasen} Hasenfratz P, Maggiore M and Niedermayer M 1990 Phys. Lett. B245 522;\\
 Hasenfratz P and Niedermayer F 1990 Phys. Lett. B245 529.
\bibitem{yam2} Yamada H 2012 Braz. J. of Phys. 42 445  (arXiv: 1209.3396 [hep-lat]).

\bibitem{cara} Caracciolo S, Edwards R G, Mendes T, Pelissetto A and Sokal A D 1995 Nucl. Phys. B (Proc. Suppl.) 47 763.
\bibitem{all} All\'es B, Buonanno A  and Cella G 1997 Nucl. Phys. B500 513.

\end{thebibliography}
\end{document}